\documentclass[iop,revtex4]{emulateapj}

\newcommand{\mytilde}{\raise.19ex\hbox{$\scriptstyle\sim$}}
\newcommand{\ciza}{CIZA~J2242.8+5301}
\newcommand{\tb}{RX~J0603.3+4214}

\usepackage{graphicx}

\shorttitle{DARK MATTER DISTRIBUTION OF \tb}
\shortauthors{Jee et al.}

\begin{document}

\title{MC$^2$: MAPPING THE DARK MATTER DISTRIBUTION OF THE ``TOOTHBRUSH" CLUSTER \tb~WITH {\it HUBBLE SPACE TELESCOPE} AND SUBARU WEAK LENSING$^*$}

\altaffiltext{*}{Based on observations made with the NASA/ESA {\it
Hubble Space
Telescope},
obtained at the Space Telescope Science Institute, which is operated by the Association of
Universities for
Research in Astronomy, Inc.}

\author{M.~JAMES~JEE\altaffilmark{1,2},  WILLIAM A. DAWSON\altaffilmark{3}, ANDRA STROE\altaffilmark{4}, , DAVID WITTMAN\altaffilmark{2}, REINOUT J. van WEEREN\altaffilmark{5}, MARCUS BR{\"U}GGEN\altaffilmark{6},  MARU{\v S}A BRADA{\v C}\altaffilmark{2}, AND HUUB R{\"O}TTGERING\altaffilmark{4}  }

\altaffiltext{1}{Department of Astronomy and Center for Galaxy Evolution Research, Yonsei University, 50 Yonsei-ro, Seoul 03722, Korea}
\altaffiltext{2}{Department of Physics, University of California, Davis, One Shields Avenue, Davis, CA 95616, USA}
\altaffiltext{3}{Lawrence Livermore National Laboratory, P.O. Box 808 L-210, Livermore, CA, 94551, USA}

\altaffiltext{4}{Leiden Observatory, Leiden University, P.O. Box 9513, NL-2300 RA Leiden, The Netherlands}

\altaffiltext{5}{Harvard-Smithsonian Center for Astrophysics, 60 Garden Street, Cambridge, MA 02138, USA}
\altaffiltext{6}{Hamburger Sternwarte, Gojenbergsweg 112, D-21029 Hamburg, Germany}

\begin{abstract}
The galaxy cluster \tb~ at $z=0.225$ is one of the rarest clusters boasting an extremely large ($\mytilde2$~Mpc) radio relic. Because of the remarkable morphology of the relic, the cluster is nicknamed the ``Toothbrush Cluster". Although the cluster's underlying mass distribution is one of the critical pieces of information needed to reconstruct the merger scenario responsible for the puzzling radio relic morphology, its proximity to the Galactic plane $b\sim10\degr$ has imposed significant observational challenges. We present a high-resolution weak-lensing study of the cluster with Subaru/Suprime Cam and {\it Hubble Space Telescope} imaging data. Our mass reconstruction reveals that the cluster is composed of complicated dark matter substructures closely tracing the galaxy distribution, in contrast, however, with the relatively simple binary X-ray morphology. Nevertheless, we find that the cluster mass is still
dominated by the two most massive clumps aligned north-south with a $\mytilde$3:1 mass ratio ($M_{200}=6.29_{-1.62}^{+2.24}\times10^{14} M_{\sun}$ and $1.98_{-0.74}^{+1.24}\times10^{14} M_{\sun}$ for the northern and southern clumps, respectively).
The southern mass peak is $\mytilde2\arcmin$ offset toward the south with
respect to the corresponding
X-ray peak, which has a ``bullet"-like morphology pointing south.
Comparison
of the current weak-lensing result with the X-ray, galaxy, and
radio relic suggests that perhaps the dominant mechanism responsible for
the observed relic may be a high-speed collision of the two most massive subclusters, although the peculiarity of the morphology
necessitates involvement of additional subclusters.
Careful numerical simulations should follow in order to
obtain more complete understanding of the merger scenario
utilizing all existing observations.
\end{abstract}

\keywords{
gravitational lensing: weak ---
dark matter ---
cosmology: observations ---
X-rays: galaxies: clusters ---
galaxies: clusters: individual (\tb) ---
galaxies: high-redshift}

\section{INTRODUCTION} \label{section_introduction}

In the hierarchical structure formation paradigm, merging is among the dominant mechanisms by which galaxy clusters grow. 
Therefore, detailed studies of merging clusters shed light on
the growth of cosmological structures.
Apart from cosmological interests, merging clusters are also 
receiving growing attention as astrophysical laboratories, providing rare and invaluable opportunities to investigate the origin of cosmic rays (e.g., Volk et al. 1996; Berezinsky et al. 1997; Feretti et al. 2012; Brunetti 
\& Jones 2014), generation of nonthermal energy in plasma (e.g., Cassano 
\& Brunetti 2005), properties of dark matter (Kahlhoefer et al.
2014), star formation and galaxy evolution driven by merging (e.g., Stroe et al. 2015), etc.. 

``Radio relic" clusters are a subclass of merging clusters that exhibit elongated diffuse radio emissions at the periphery of the systems.
These ``radio relics" often occur in pairs and in most cases stretch nearly perpendicular to observed merger axes. Now many observational and theoretical studies support
the premise that the relics trace the locations of shock fronts induced by cluster mergers (e.g., Ensslin et al. 1998). 
Detailed analysis of the radio relic data enables us to put  independent constraints on the key parameters necessary in our reconstruction of the merging scenario,
including the direction of the merger, the projection angle between the merger axis and the plane of the sky, the shock velocity, and the time
since the impact (e.g., Ng et al. 2015).
Because of the limited observational time window set
by both development and deterioration of mature shocks, only a few tens of radio relic clusters are known to date (e.g., van Weeren et al. 2010; 2012; 2013, Govoni et al. 2001, Brunetti et al. 2008).

\begin{figure}
\centering
\includegraphics[width=8.5cm,trim=2.8cm 0cm 2.5cm 0cm]{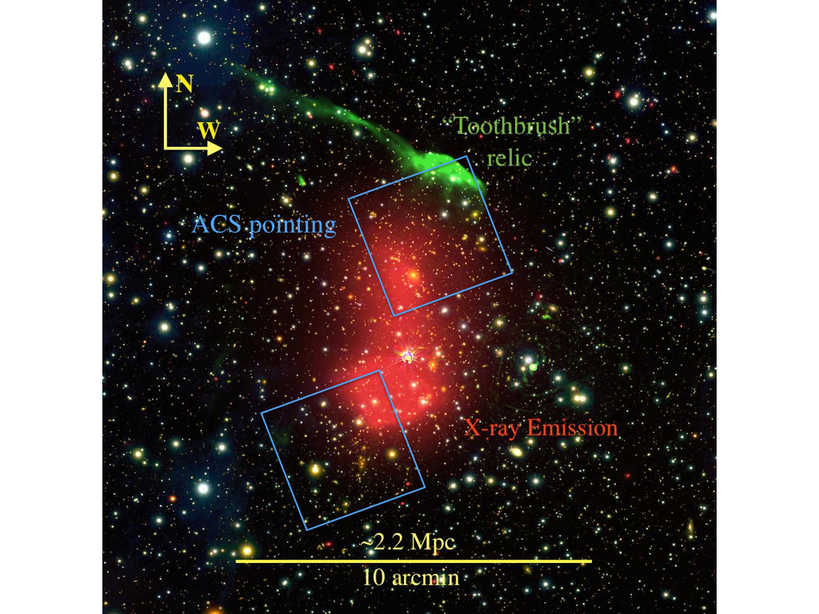}
\vspace{0cm}
\caption{Illustration of different cluster components in the merging cluster \tb. The  intensity in green represents the 610 MHz radio emission measured with GMRT (van Weeren et al. 2012). The intensity in red shows the X-ray emission observed with Chandra. The background color composite is created using Subaru/Suprime Cam data with the $g$, $r$, and $i$ filters depicting the intensity in blue, green, and red channels, respectively. We subtracted the bright ($m_i\sim7$) star located at the center.
The two blue rhombuses depict the two pointings and orientations of Advanced Camera for Surveys (ACS).
}
\label{fig_tb}
\end{figure}

The cluster \tb~is a remarkable cluster at $z=0.225$ whose radio relic stretches over $\mytilde2$ Mpc (Figure~\ref{fig_tb}). 
Because of its peculiar morphology composed of the western short ($\mytilde0.5$ Mpc) thick band (``brush")
and a long ($\mytilde1.5$ Mpc) thin stripe (``handle"), \tb~is nicknamed the ``Toothbrush Cluster" (van Weeren et al. 2012).
Together with the much fainter relic found near the southern cluster edge,
this asymmetric and remarkably linear feature implies that perhaps the merger might have been complex, involving more than two subclusters (Bruggen et al. 2012).
This unusual radio morphology is different from that of the \ciza~ cluster (van Weeren et al. 2010), possessing a similarly giant but more symmetric ``sausage"-like radio relic.

The ``Toothbrush" relic of \tb~was discovered with the Westerbork Synthesis Radio Telescope (WSRT) and the Giant Metrewave Radio Telescope (GMRT) by van Weeren et al. (2012). 
For the ``Toothbrush" relic, van Weeren et al. (2012) detected a 
spectral index gradient from the front (northern edge) of the ``Toothbrush" relic toward the back (southern edge). The frontal part of this relic is  highly polarized ($\mytilde60$\% at 4.9~Ghz),
which indicates that the merger might be happening nearly in the plane of the sky (Ensslin et al. 1998).

Van Weeren et al. (2012) also found a north-south elongated X-ray morphology at the location of the cluster based
on archival ROSAT data.
Bruggen et al. (2012) carried out a numerical simulation of \tb~
by modeling the cluster with two large ($5\times10^{14}M_{\sun}$)  and one small ($3.5\times10^{13}M_{\sun}$) halos and demonstrated
that the simulation can generate a giant relic with a similar morphology.
Ogrean et al. (2013) studied the cluster with {\it XMM-Newton} data, which reveal two distinct X-ray peaks. At both northern and southern edges (near the relics) of the cluster, they detected density discontinuities indicating the presence of potential shocks. 
Itahana et al. (2015) constrained the Mach number to be $\mathcal{M}\sim1.6$ based on this density discontinuity, which is consistent with their independent measurement $\mathcal{M}\sim1.5$ from the temperature
jump obtained from {\it Suzaku} data. However, these X-ray-based Mach numbers are lower than what the radio data imply ($\mathcal{M}\sim4$; van Weeren et al. 2012).
Correlations between the cluster galaxy star formation and the merger environment were studied by Stroe et al. (2014; 2015).

Despite a number of studies mentioned above on this remarkable system,
no reliable mass estimation of the system has been carried out, and little is known about the spatial distribution of its mass and member galaxies.
The cluster's underlying mass distribution is one of the critical pieces of information in order to infer the merger scenario responsible for the radio relic morphology (e.g., Ng et al. 2015; Dawson 2013).
Hence, in this paper, as part of our Merging Cluster Collaboration\footnote{http://www.mergingclustercollaboration.org/} (MC$^2$) project, we present detailed weak-lensing analysis of \tb~ with Subaru/Suprime Cam and {\it Hubble Space Telescope (HST) } imaging data. Because of the low galactic latitude $b\sim10\degr$
of the system, some observational challenges, including severe extinction and stellar obscuration, are present. However, in the weak-lensing study of \ciza~(Jee et al. 2015), we already demonstrated that a successful weak-lensing study is still possible when high-resolution imaging observations are carefully planned and analyzed with the state-of-the-art technique.

We launched the MC$^2$ project to study a large sample of merging clusters with a coherent approach.
Our immediate goals for the current paper are (1) to map the underlying mass distribution and compare the result with the galaxies and X-ray emission and (2) to quantify the matter content of the system. These mass properties are among the critical parameters necessary to constrain the merging scenario of the system leading to
such an unusual morphology in radio emission. Our long-term goals of the MC$^2$ project include detailed studies of dark matter properties through systematic investigation of the large sample and careful numerical simulations.

We present our study as follows. In \textsection\ref{section_observation} we describe our data and reduction. In \textsection\ref{section_wl} we review the basic lensing theory of weak lensing and our technique.
We present our mass reconstruction results in \textsection\ref{section_mass_reconstruction}, discussing
the source selection, mass distribution, and mass estimation.
Discussions of our results will follow in \textsection\ref{section_discussion} before
we conclude in \textsection\ref{section_conclusion}.

We assume a flat $\Lambda$ cold dark matter ($\Lambda$CDM) cosmology with $H_0=70~\mbox{km}~\mbox{s}^{-1}~\mbox{Mpc}^{-1}$, $\Omega_M=0.3$, and $\Omega_{\Lambda}=0.7$. At the redshift of \tb, $z\sim0.225$, the plate scale is $\mytilde 3.61~\mbox{kpc}/\arcsec$ ($\mytilde217~\mbox{kpc}/\arcmin$). The $M_{200}$ value that we adopt here as a halo mass is a spherical mass within $r_{200}$, inside which the mean density becomes 200 times the critical density of the Universe at the redshift of the cluster.

\section{OBSERVATIONS AND DATA REDUCTION} \label{section_observation}

\subsection{Subaru/Suprime Cam}

\tb~ was observed with Subaru/Suprime Cam on 2013 February 25 in $g$, $r$, and $i$ with total integrations of 720, 2880, and 720~s, respectively.
We used four visits for $g$ and $i$ and eight visits for $r$ with varying roll angles in order to remove cosmic rays and mitigate the impact of ``bleeding" trails while co-adding. 
As demonstrated in our analysis of \ciza~(Jee et al. 2015), this rotation of fields significantly increases the number of usable galaxies for weak-lensing
study of clusters at low galactic latitude, whose shapes otherwise would have been affected by a number of 
``bleeding" trails and diffraction spikes.

The low-level CCD processing (overscan subtraction, bias correction, flat-fielding, initial geometric distortion rectification, etc.) was carried out with the SDFRED2 
package\footnote{http://subarutelescope.org/Observing/Instruments/SCam/sdfred}. 
We refined the geometric distortion and World Coordinate System (WCS) information using the SCAMP software (Bertin 2006)\footnote{http://www.astromatic.net/software/scamp}. The Two Micron All Sky Survey (Skrutskie et al. 2006) catalog was
selected as a reference when the SCAMP software was run. We also rely on SCAMP to calibrate out the sensitivity variations across different frames.
For image combination, we ran the SWARP software (Bertin et al. 2002)\footnote{http://www.astromatic.net/software/swarp} using the SCAMP result as inputs.
We first created a median mosaic image and then used it to mask out pixels ($3\sigma$ outliers) in individual frames. These masked individual frames were weight-averaged to generate the final mosaic image, which we use for our scientific analysis presented hereafter.

A very bright ($m_i\sim7$) star was located at the approximate cluster center (i.e., between the two X-ray peaks) of \tb~(R.A.,decl.)=(6:03:17.5, 42:12:25), and from visual inspection we find that
its halo is affecting a substantial area (a circular region with $d\sim5\arcmin$). In order to investigate the impact of this $m_i\sim7$ star on our mass reconstruction, we 
experimented with two star-subtraction schemes. In the first method, we let SWARP determine the local sky level (thus effectively the surface brightness level of the point-spread-function [PSF] wing near the bright star) and remove it.
This method removes a substantial amount of the sky gradient and allows us to detect many galaxies within the PSF wing. However, one notable weakness of this method is that many high-frequency features remain after the subtraction because the spatial resolution of the sky estimation by SWARP is limited.
In the second method, we preserve the sky level in our image reduction and only subtract the PSF from the final co-add after modeling the PSF profile.
Judging from visual inspection, we believe that the second method is superior in terms of high-frequency feature removal.
Nevertheless, we find that our weak-lensing results from both image reduction schemes are highly consistent not only in total mass estimation but also in spatial
mass reconstruction. The test reassures us that our analysis is robust against the details of the central stellar light subtraction.

For object detection and shape catalog generation, we refer readers to Jee et al. (2015).
In brief, we run SExtractor (Bertin \& Arnouts 1996) in a 
dual-image mode using the $r$-band image for detection.
The blending threshold parameter
({\tt BLEND\_NTHRESH}) is set to 32 with a minimal contrast of
{\tt DEBLEND\_MINCONT}=$10^{-4}$.
We employ redenning values
from Schlafly \& Finkbeiner (2011) to correct for dust attenuation.
We measure object shapes from the $r$-band images, which provides $\mytilde0.7\arcsec$ seeing on average.

\subsection{Hubble Space Telescope}
The two optically densest regions (Figure~\ref{fig_tb}) of \tb~were observed with {\it HST} using both Advanced Camera for Surveys (ACS) and Wide Field Camera 3 (WFC3) in parallel during the 2013 October 10 and 2014 January 24 periods under the program HST-GO-13343.
The distance
between the two instruments on the projected plane of sky is approximately 6$\arcmin$, which fortuitously corresponds to the separation between the two regions. Each region was
imaged with two orbits of ACS F814W, one orbit of WFC3 F606W, and one orbit of WFC3 F390W. 

Charge transfer inefficiency (CTI) is an important issue when dealing with CCDs in space as high-energy particles damage the detectors and 
create a growing number of traps.
The effect is severe
in both detectors, which, if  uncorrected for, would leave
substantial ``charge trails" and compromise our scientific capability.
The current pipeline of the STScI automatically corrects for the effect using the latest pixel-based method 
(Ubeda \& Anderson 2012), however only for ACS. Thus, for the WFC3 data, we manually applied
the preliminary version of the STScI script {\tt wfc3uv\_ctereverse\_parallel.F} to raw data to correct for the CTI effect.
The importance of accurate CTI correction for {\it HST} weak-lensing analysis is discussed in the study of Jee et al. (2014), which concludes that the automatic CTI correction for ACS data by the STScI pipeline 
is adequate for cluster weak lensing, although the method tends to overcorrect the effect at the faint limit.

The software MultiDrizzle (Koekemoer et al. 2002) is used to rectify detector distortions, remove cosmic rays, and create stacks. A critical input to MultiDrizzle is the information regarding accurate relative offsets between images. Within each visit, the typical shift is less than a pixel. However, for different visits, the shift can become as large as a few tens of pixels.
We used common astronomical objects to measure relative offsets. 
The estimated alignment error is $\mytilde0.01$ pixels, which easily meets the
cluster weak-lensing requirement.
We ``drizzle" images with the final pixel scale of 0.05 $\arcsec\mbox{pixel}^{-1}$ and the Lanczos3 kernel. Readers are referred to our previous paper (e.g., Jee et al. 2014)
for more details regarding the HST data reduction in the context of weak lensing.
We measure object shapes only from the ACS F814W images for weak-lensing analysis, although the F606W-F814W colors are used to identify the cluster members.

\subsection{Keck DEIMOS Spectroscopic Observation}
Detailed description of our DEIMOS spectroscopic observation and data reductions is provided by Dawson et al. (2015). Here we only present a brief summary.
We carried out  a spectroscopic survey of \tb~with the DEIMOS instrument during two observing runs on 2013 January 16 and September 5  
using $1\arcsec$ wide slits with the 1200 line mm$^{-1}$ grating. The resulting pixel scale is 0.33 \AA~pixel$^{-1}$ and a resolution of $\mytilde$1 \AA~ (50 km s$^{-1}$).
We obtained a total of 419 spectra, of which we were able to determine reliable redshifts for 390 objects. We define 240 spectroscopic galaxies 
within the range 0.21$<z<0.24$ as cluster members.

\section{WEAK-LENSING METHOD} \label{section_wl}
Although accurate measurement of subtle shape distortions of galaxy images by overcoming various sources of instrumental systematic effects is technically nontrivial, studies of galaxy clusters with weak lensing have been firmly established as powerful methods to investigate the mass and its distribution. Readers are referred to many excellent reviews in the literature for a more complete description of the technique and issues (e.g., Bartelmann \& Schneider 2001). Here we provide a summary of the theory and the data analysis method.

\subsection{Theoretical Background}
The coordinate transformation by gravitational lensing in a weak-lensing regime is often expressed by the following matrix $\textbf{A}$:
\begin{equation}
\textbf{A}=(1-\kappa) \left (\begin{array} {c c} 1 - g _1 & -g _2 \\
                      -g_2 & 1+g _1
          \end{array}  \right ), \label{eqn_lens_trans}
\end{equation} 
\noindent
where $\kappa$ and $g_{1(2)}$ are convergence and reduced shears, respectively. 
A positive value of $g_1$ stretches the shape of an object in the $x$-axis direction, whereas a negative value elongates the object in the $y$-axis direction. Similarly, a positive value of $g_2$ is responsible for the elongation along the direction defined by the function $y=x$ (i.e., 45$\degr$ with respect to the $x$-axis.
We refer to $g=(g_1^2 + g_2^2)^{1/2}$ as a ``reduced" shear in order to distinguish it from a shear $\gamma$:
\begin{equation}
\gamma = (1-\kappa) g.
\end{equation}
\noindent
$\kappa$ is the projected mass density expressed in units of the critical surface mass density:
\begin{equation}
\Sigma_c = \frac{c^2}{4 \pi G D_l \beta } \label{eqn_sigma_c}.
\end{equation}
\noindent
In Equation~\ref{eqn_sigma_c}, $c$ is the speed of light, $G$ is the gravitational constant,
and $D_l$ is the angular diameter distance to the lens. $\beta$ is the angular diameter distance ratio defined as $D_{ls}/D_s$, where
$D_{ls}$ and $D_s$  are the angular diameter
distances between the lens and the source and between the observer
and the source, respectively. In typical weak-lensing studies, accurate redshifts of individual galaxies are unknown, and thus it is common to estimate $\beta$
for the entire source population, which inevitably contains some foreground galaxies.
In this case, $\beta$ is given as
\begin{equation}
\beta=max \left [ D_{ls}/D_s,0 \right ]. \label{eqn_beta}
\end{equation}
Because the lensing kernel is nonlinear, using the effective mean value $\beta$ above biases the result. An analytic first-order correction is derived by Seitz \&  Schneider (1997), and we apply the method to our analysis.

\subsection{Implementation: Shape Measurement and PSF Modeling}
The matrix (eqn.~\ref{eqn_lens_trans}) transforms a circle into an ellipse, and the resulting ellipticity becomes $g$ when we define ellipticity as $e=(a-b)/(a+b)$, where $a$ and $b$ are the semimajor and semiminor axes, respectively. Therefore, when no bias is present, the measurement of $g$ is simply averaging the object's (ideal) ellipticities. That is,
\begin{equation}
g_{1(2)} = \left < e_{1(2)} \right > \label{eqn_e_ave}
\end{equation} 
\noindent where $e_1$ and $e_2$ are computed by measuring $a$, $b$, and $\theta$ (angle between the semi-major and the positive $x$-axes) as follows: 
\begin{equation}
e_1= e \cos(2 \theta)  \label{eqn_e1}
\end{equation}
\begin{equation}
e_2= e \sin(2 \theta). \label{eqn_e2}
\end{equation}
\noindent

Now the important question is how one measures ellipticity from observed galaxy images, which are not only complex but also subject to distortions from nongravitational lensing sources such as atmospheric and optical aberrations, detector anomalies, and image processing artifacts. Extensive discussions on these challenges are available in the literature, and for some challenges many state-of-the-art algorithms meet or exceed the requirements that many future weak-lensing surveys demand (Mandelbaum et al. 2015). Below we briefly describe our shape measurement method, which
turns out to be among the best-performing methods in the 3rd GRavitational lEnsing Accuracy
Testing (GREAT3; Mandelbaum et al. 2014; 2015). The GREAT3 challenges include realistic PSFs and their spatial variations, realistic galaxy morphologies, multi-epoch data, etc.

We model an observed (smeared by PSF) galaxy image with a convolution of an elliptical Gaussian function $G(x,y)$ and a PSF $P(x,y)$:
\begin{equation}
M (x,y) = G (x,y) \otimes  P(x,y).
\end{equation}
The elliptical Gaussian function $G(x,y)$ has four parameters, namely, the semimajor axis $a^{\prime}$, the semiminor axis $b^{\prime}$, the position angle $\theta$, and the normalization $n$; we let the centroid remain fixed. $P(x,y)$ is computed by applying principal component analysis (PCA) to stellar images. 

Our PSF modeling scheme is different between Subaru and {\it HST}. For Subaru, each CCD frame contains a sufficient number ($\gtrsim100$) of high S/N ($>20\sigma$) stars, which enables us to apply PCA directly to science images (Jee \& Tyson 2011). This is not the case for HST, whose small field of view ($\mytilde3\arcmin\times\mytilde3\arcmin$) normally provides only 10-20 high S/N stars\footnote{For the current target, the stellar density
is a few times higher than this average value because of its proximity to the Galactic plane. 
However, this still does not allow us to obtain reliable PSF models based on these stars in the science image.}
Therefore, we use external stellar field images and construct PSF libraries from them (Jee et al. 2007a).
Of course, it is necessary to find a matching PSF template from the library for each science frame, where we measure weak lensing. This template matching between science and stellar fields is possible because the {\it HST} PSF pattern is repeatable, largely determined by the focus
as empirically demonstrated by Jee et al. (2007a).
Schrabback et al. (2010) suggest that perhaps two parameters might be needed to better characterize
the PSF pattern of a given ACS exposure.

Going back to the issue of ellipticity measurement, we minimize the difference between $M(x,y)$ and the observed galaxy profile $O(x,y)$. We refer to the ellipticity from this measurement as raw ellipticity $e^{\prime}$. This raw ellipticity $e^{\prime}$
is slightly offset from the ideal ellipticity $e$ above, which we convert to the reduced shear by straightforward averaging.
The sources of the bias include noise bias, model bias, truncation bias, etc (see Mandelbaum et al. 2015 for details). 
Thus, we modify the above Equation~\ref{eqn_e_ave} as follows:
\begin{equation}
g_{1(2)} = m_{1(2)} \frac{1}{W} \sum_{i=1}^{N} e_{1(2)}^{\prime} \mu_i
\end{equation}
\noindent
where $\mu_i$ is the inverse-variance weight,
\begin{equation}
\mu_i  = \frac {1} { \sigma_{SN}^2 + (\delta e_i)^2}, \label{eqn_shear_weight}
\end{equation}
\noindent
$W$ is $W=\Sigma \mu_i$, and $m_{1(2)}$ is the multiplicative bias, which is empirically determined from our image simulations (e.g., Jee \& Tyson 2011; Jee et al. 2013). The simulated images match the depth, resolution, and source density of our science images, which are
necessary to capture the dependence on imperfect deblending, PSF size, noise level, etc.
Although we find that the multiplicative bias $m_{1(2)}$ is a function of several parameters such as S/N, size, and surface brightness, we
derive a single value that represents the mean correction for the entire source population. 
For HST and Subaru data, we obtain $m_{1}=m_{2}=1.09$
and $m_{1}=m_{2}=1.14$, respectively; 
in the derivation of these  calibration factors, we also apply the same source selection criteria discussed in \textsection\ref{section_source_selection}.
Apart from this multiplicative bias, additive bias is
present especially when PSFs are severely elongated. We find that the mean level of additive bias is a few percent of
the PSF ellipticity. This level of bias is certainly a concern for cosmic shear measurements (Jee et al. 2013) but can be safely ignored in 
the current cluster lensing.
In equation~\ref{eqn_shear_weight}, $\sigma_{SN}$ is the ellipticity dispersion per component measured from the data. $\delta e_i$ is the ellipticity measurement noise per component, which is derived from the Hessian matrix (obtained from the likelihood function for elliptical Gaussian fitting). 

\begin{deluxetable*}{lcccccc}
\tabletypesize{\scriptsize}
\tablecaption{Statistical properties of source galaxies in different regions.}
\tablenum{1}
\tablehead{\colhead{Available Data} &  \colhead{Area (arcmin$^2$)} & \colhead{Density (arcmin$^{-2}$)} &
\colhead{Shape}   & \colhead{Color} & \colhead{$<\beta>$} &  \colhead{$<\beta^2>$}  }
\tablewidth{0pt}
\startdata
Subaru + ACS + WFC3   & 14.6  & 59 & ACS F814W   & F606W-F814W & 0.697         & 0.503 \\
Subaru + ACS          & 11.0  & 66 & ACS F814W   & $g-r$       & 0.680         & 0.481 \\
Subaru                & 384   & 31 & Subaru $r$  & $g-r$     & 0.672         & 0.495 
\enddata
\end{deluxetable*}

\section{WEAK-LENSING RESULTS} \label{section_mass_reconstruction}

\subsection{Source Selection and Redshift Estimation \label{section_source_selection}}
\begin{figure}
\includegraphics[width=9cm]{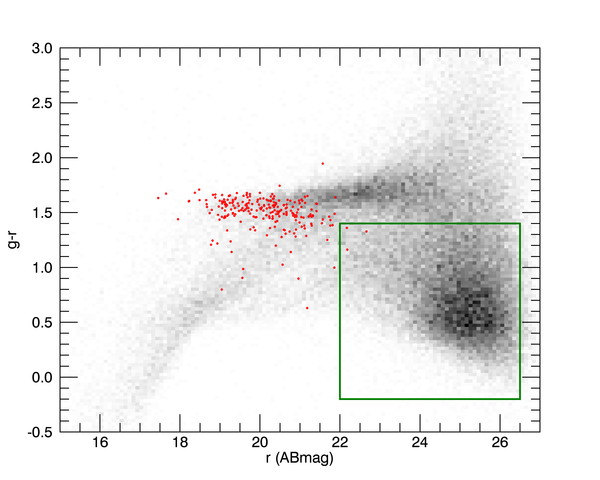}
\caption{Subaru color-magnitude relation in the \tb~field. We observe a tight color-magnitude relation of the red-sequence galaxies.
The red circles are spectroscopically confirmed cluster members. 
The green box represents our selection function for the source population.} \label{fig_cmd}
\end{figure}

Following Jee et al. (2015), we rely on the color-magnitude relation to select cluster members and lensing sources. The so-called ``4000\AA~break" redshifted to the cluster at $z=0.225$ is well bracketed by the $g-r$ color, and it is straightforward to identify the red sequence of \tb~with the color-magnitude relation.
Figure~\ref{fig_cmd} shows that
this relation appears to continue down to $m_r\sim23$, where the red-sequence tail starts to blend into the faint ``cloud".
We construct a  cluster member catalog by combining our spectroscopically confirmed $\mytilde240$ members (W. Dawson et al. in prep.) and the red sequence defined by Subaru and HST photometry.
For the Subaru cluster member catalog, we select sources whose $(g-r)$ colors are between 1.4 and 1.8 and whose $r$-band magnitude is brighter than 23. 
In the region where ACS colors are available, fainter cluster members (F814W$\lesssim25$) are selected.
We combine the two catalogs and remove spectroscopically confirmed nonmembers. 
In \textsection\ref{section_massmap}, we use the smoothed luminosity map created from this catalog for comparison with the mass distribution.

We define the source population as the objects with colors bluer than the red sequence. Our selection criteria are
\begin{eqnarray}
& -0.2  <  g-r  < 1.4 \nonumber \\
& 22  <  r  < 26.5
\end{eqnarray}
for Subaru and
\begin{eqnarray}
& -0.3<F606W-F814W<1.49  \nonumber \\
& 20<F814W<27 
\end{eqnarray}
for {\it HST}.
In addition to the above color and magnitude cuts, we also apply shape criteria cuts. Namely,
the post-seeing half-light radius $r_h$ should be greater than the value for stars, the shape measurement error $\delta e$ should be less than 0.3, and the pre-seeing semiminor axis $b$ should be greater than 0.3 pixels. For the region where only the ACS F814W filter is available, we use Subaru
colors. The mean number density of sources is $\mytilde31~\mbox{arcmin}^{-2}$ in the Subaru-only region, whereas the source density becomes a factor of two higher (59--66$~\mbox{arcmin}^{-2}$) in the ACS region.

We experimented with several different selection criteria by altering color and magnitude cutoffs by up to $\mytilde0.6$ and found that the resulting mass reconstruction (\textsection\ref{section_massmap}) is not sensitive to the choice as long as the final source number densities are within $\mytilde20$\% of the values above. This indicates that the background fraction does not vary significantly when we perturb the criteria in
the neighborhood of the above selection.

In order to obtain the redshift distribution of our source population, we use the photometric
redshift catalog of Dahlen et al. (2010)
from the Great Observatories
Origins Deep Survey (GOODS; Giavalisco et al. 2004) data. GOODS consists of two separate fields, GOODS-N and GOODS-S, each   covering $\mytilde160~\mbox{arcmin}^2$. We combine both photometric
redshift catalogs. 

For the purpose of source redshift determination, our weak-lensing field can be divided into the following three regions, where the available data are (1) only Subaru shapes and Subaru colors, (2) HST shapes and HST colors, and (3) HST shapes and Subaru colors.

For the first case, we perform photometric transformation of the $g-r$ color
to match the ACS colors. We obtain $\beta=0.672$ (eqn.~\ref{eqn_beta}) after taking into account the difference in depth; without this depth correction, a slightly higher value of $\beta=0.705$ is estimated.
The width of the distribution should also be determined to correct for the bias arising from the assumption that all sources lie at the single redshift plane.
We measure $\left <\beta^2 \right >$ to be 0.495.
For the second case, we assume that our F814W in \tb~matches F775W in GOODS.
With this assumption, $\left <\beta \right >$ and $\left <\beta^2 \right >$ are estimated to be 0.697 and 0.593, respectively.
Finally, for the final case, we obtain $\left <\beta \right>=0.680$ and $\left <\beta^2 \right>=0.481$. These source redshift characteristics are summarized in Table 1.

\subsection{One-dimensional Analysis} \label{section_1D}

A traditional method of representing weak-lensing signals is a reduced tangential shear profile. This provides a measure of how strongly source galaxies are tangentially aligned around a reference point, often chosen to be the center of a cluster. The mathematical definition of the reduced tangential shear is given as
\begin{equation}
 g_T  = -  g_1 \cos 2\phi - g_2 \sin 2\phi \label{tan_shear},
\end{equation}
\noindent
where $\phi$ is the position angle of the object with respect to the reference axis.

\begin{figure}
\includegraphics[width=8.5cm,trim=0cm 0cm 0cm 0cm]{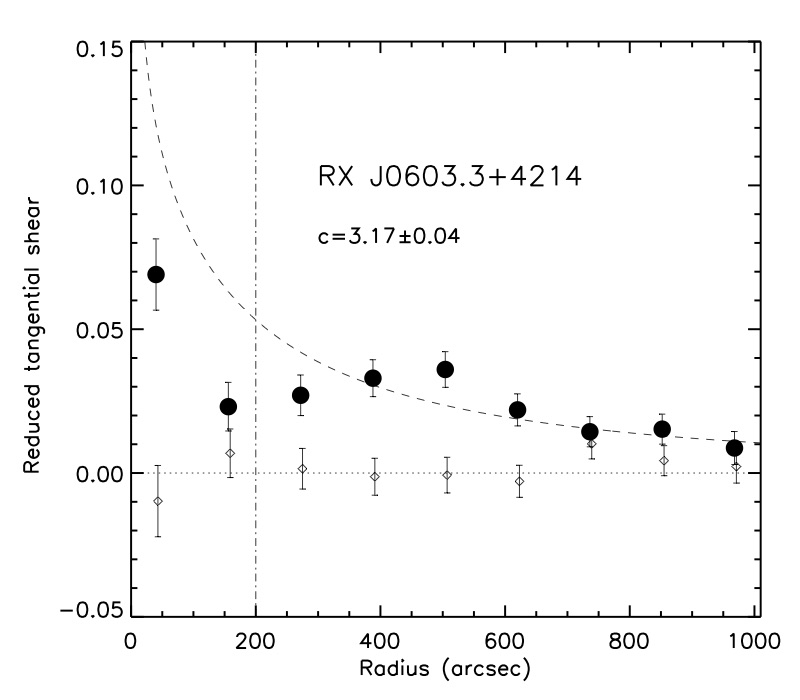}
\caption{Reduced tangential shear profile of \tb. Filled circles are reduced tangential shears azimuthally averaged with respect to the northern halo. Diamond symbols represent the results when galaxies are rotated by 45$\degr$. These results are often referred to as ``B-mode" signals and must be consistent with zero as observed when no systematics are present.
The errors are estimated by combining both ellipticity dispersion (from the data) and measurement errors.
The dashed line is the best-fit tangential shear ($c=3.17\pm0.04$) when the results at $r>200\arcsec$ are used. It appears that the cluster substructure complicates the shape of the shear profile at small radii ($r\sim200\arcsec$).
}
\label{fig_tan_shear}
\end{figure}

The reduced tangential shear centered at the northern luminosity peak of \tb~is displayed in Figure~\ref{fig_tan_shear}; we show in \textsection\ref{section_massmap} that the 
northern halo is the strongest mass peak.
The signals are all positive within the displayed range $r<1000\arcsec$ 
($\mytilde3.6$~Mpc). Since these points are uncorrelated, the significance of the lensing detection is very high ($\mytilde10 \sigma$). Also displayed in Figure~\ref{fig_tan_shear} are so-called
B-mode signals (diamonds), which serve as a measure of residual systematics and
should be consistent with zero as observed when the systematics are
under control.
Fitting a single Navarro-Frenk-White (NFW) profile to the reduced tangential shears in merging clusters
is not a reliable method to quantify the mass. Nevertheless, this provides a quick method to
estimate the approximate total mass of the system.
Typically, because of significant substructures near the center, reduced shears at large radii are
used to estimate the global mass.
Using the data points at $r>200\arcsec$\footnote{The separation between the northern and southern
mass peaks is $\mytilde6~\arcmin$. Therefore, in principle the cutoff value $r_{min}=200\arcsec$ is not sufficiently large when one wants to minimize the bias. However, we use this 1D-fitting only to obtain a quick estimate.}
($\mytilde0.7$ Mpc) and the
mass-concentration relation of Duffy et al. (2008), we obtain $c=3.17\pm0.04$, which translates to $M_{200}=1.01_{-0.14}^{+0.16}\times10^{15}M_{\sun}$.
Our 1D analysis indicates that the total mass of the \tb~cluster is not low, but certainly not as extreme as the ones in \ciza~or El Gordo (Jee et al. 2014; 2015), whose global mass
approaches or exceeds $M_{200}\sim3\times10^{15}M_{\sun}$.

\subsection{Two-dimensional Mass Distribution} \label{section_massmap}

We present mass reconstruction results based on the maximum entropy method of Jee et al. (2007b). The method uses the ``entropy" of the mass pixels to regularize the mass map. Effectively, the resulting mass map is adaptively smoothed with a kernel depending on the local S/N. This regularization suppresses spurious features 
at the boundaries often present in old methods such as Kaiser \& Squires (1993).

\begin{figure*}
\centering
\includegraphics[width=8.5cm]{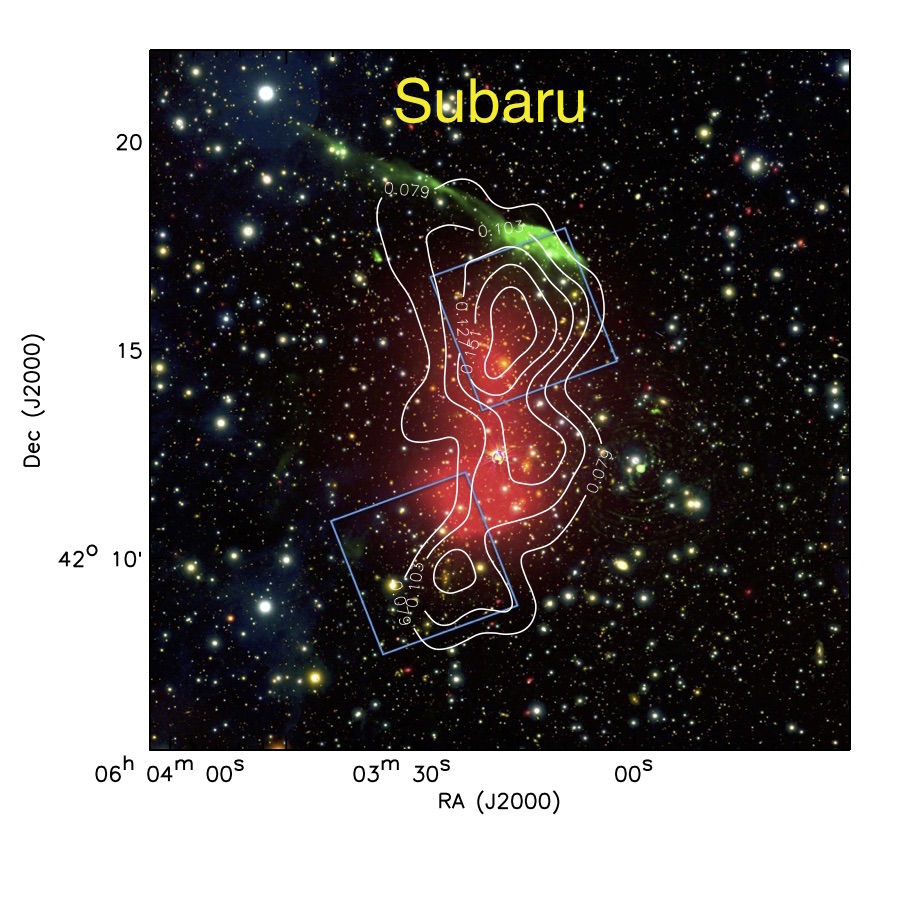}\includegraphics[width=8.5cm]{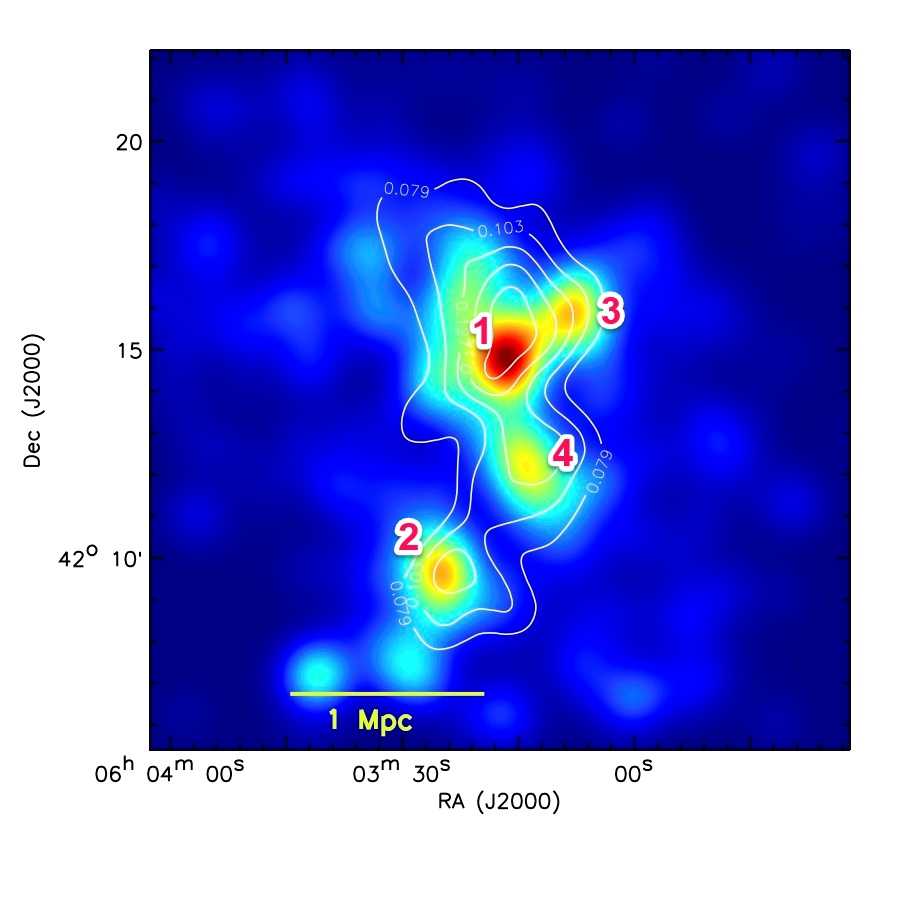}
\caption{Mass reconstruction using Subaru weak lensing. In the left panel, mass contours are overlaid on the color composite also shown in Figure~\ref{fig_tb}.
In the right
panel, we overlay mass contours on the smoothed optical ($i$-band) luminosity of the cluster members. Overall, the mass distribution follows the galaxy distribution, whereas we find a clear offset between X-ray and mass in the southern region.}
\label{fig_subaru_massmap}
\end{figure*}

In Figure~\ref{fig_subaru_massmap}, we show the mass reconstruction based on the Subaru data. The mass map clearly shows the north-south elongation seen in the distributions of the X-ray emission and cluster galaxies.
The correlation of the weak-lensing mass with the smoothed optical light ($i$ band luminosity) is high. 
The optical light is obtained from the cluster members selected based on their spectroscopic redshifts and
$g-i$ colors. The smoothing kernel is a Gaussian with FWHM$=100\arcsec$.
We identify at least four luminosity peaks, and three of them (L1, L2, and L4) are resolved
by the Subaru weak lensing. Because of the bright ($r\sim7$) star in the field center, there is a nonnegligible chance that the substructures around the star may be affected, although we carefully subtract
the stellar light profile and attempt to use as many galaxies as possible in the neighborhood.
We suspect that the influence on the mass peak near L1 is minor because of its significant lensing signal (some strong-lensing features are also visible).
However, interpretation of the substructure near L4 needs caution because it is very close to the star and also the peak significance is weaker despite the apparent alignment between light and mass.

\begin{figure*}
\centering
\includegraphics[width=8.5cm]{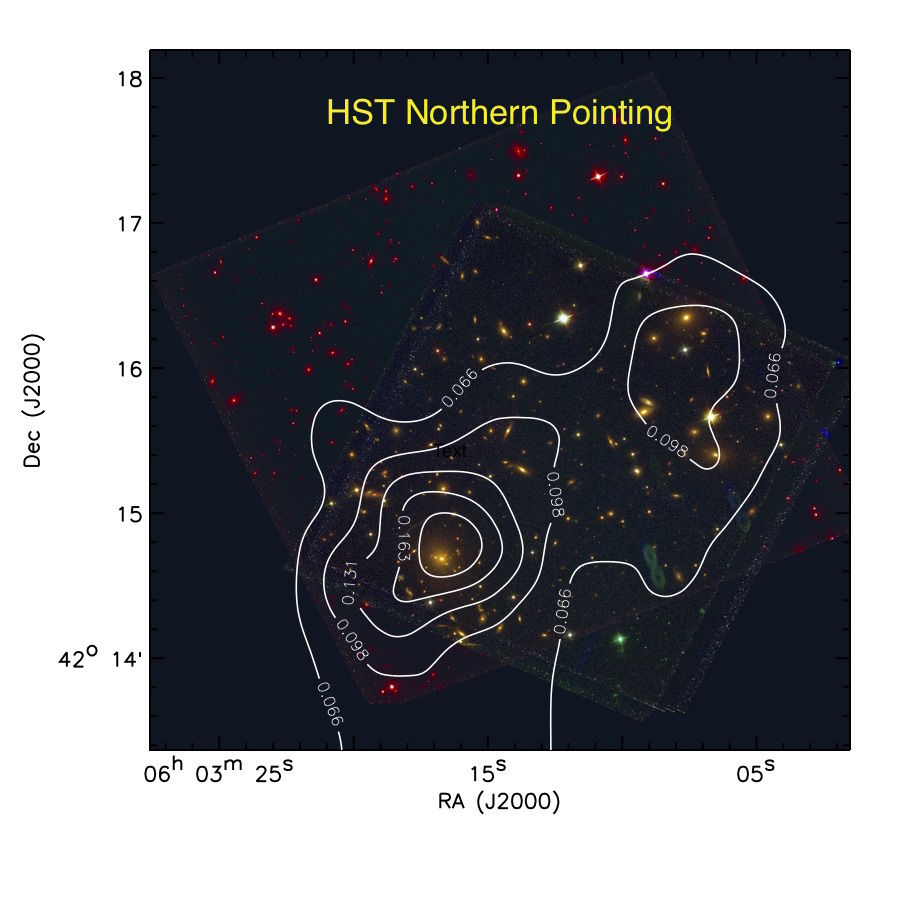}\includegraphics[width=8.5cm]{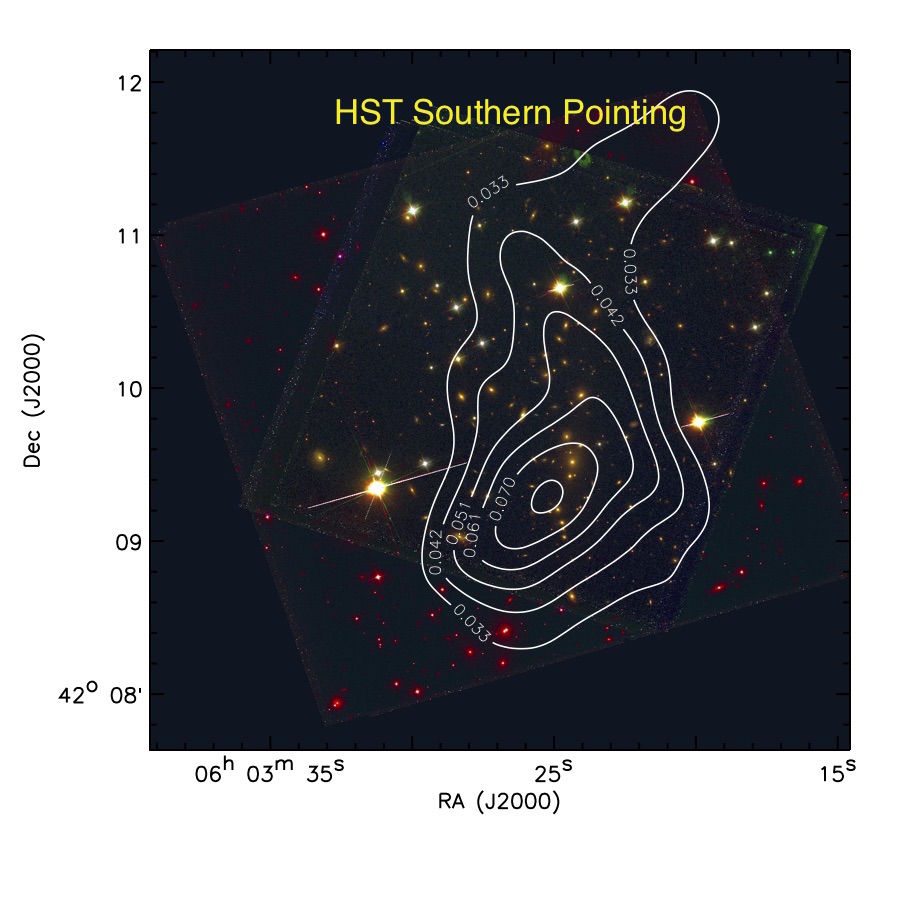}
\caption{Mass reconstruction using HST weak lensing. The color composites are created by combining the ACS F814W (red), WFC3 F606W (green), and WFC3 F390W (blue) data. Refer to Figure~\ref{fig_subaru_massmap} for guidance in locating the two HST fields within the larger Subaru fields. The northern mass map obtained from HST is consistent with the Subaru result, although the presence of more source galaxies in the former allows us to resolve the two components also traced by the cluster galaxies. The southern mass distribution also agrees nicely with the Subaru result. No distinct mass peak is found near the southern X-ray peak. However, note the extension of the HST mass map toward the peak of the X-ray emission.
}
\label{fig_hst_massmap}
\end{figure*}

We display our {\it HST} weak-lensing results in Figure~\ref{fig_hst_massmap}.
The results are consistent
with the Subaru results.
The northern mass peak in the Subaru mass map
is further resolved into two peaks thanks to the factor of two increase in the {\it HST} source density. The {\it HST} southern mass peak is in excellent agreement with the Subaru result; the two centroids are highly consistent, and both mass peaks show an extension toward the North.

\begin{figure*}
\includegraphics[width=9.cm,trim=2cm 0cm 0cm 0cm]{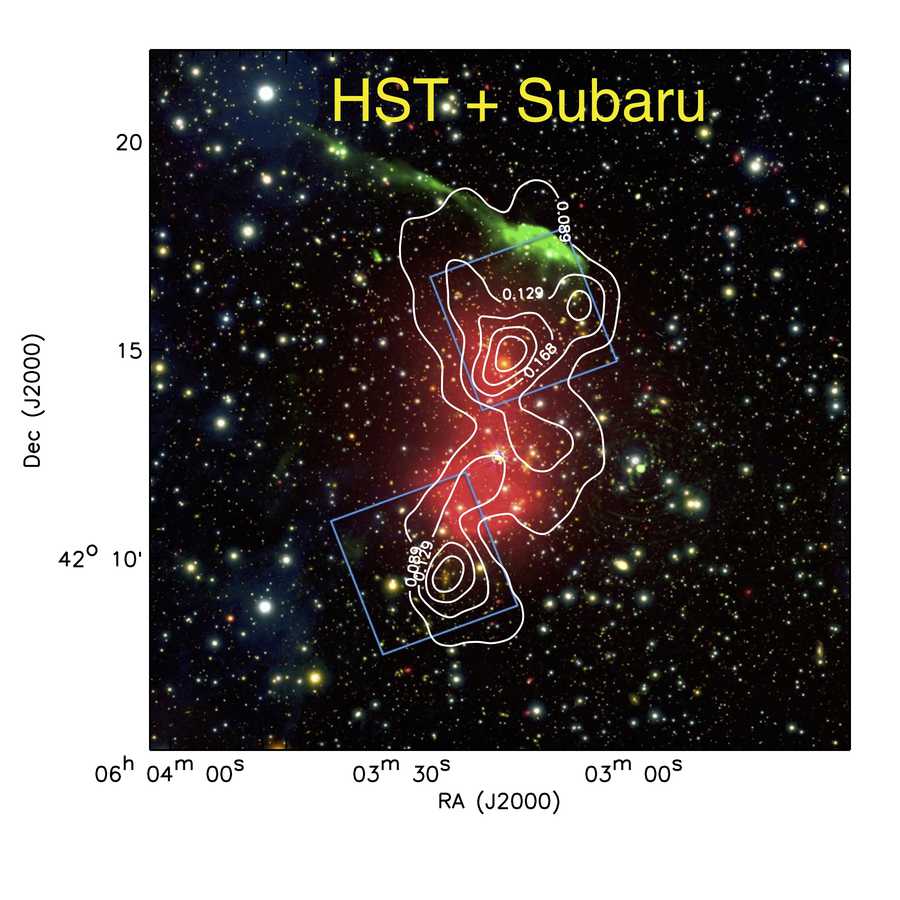}
\includegraphics[width=8.7cm,trim=2cm 0cm 0cm 0cm]{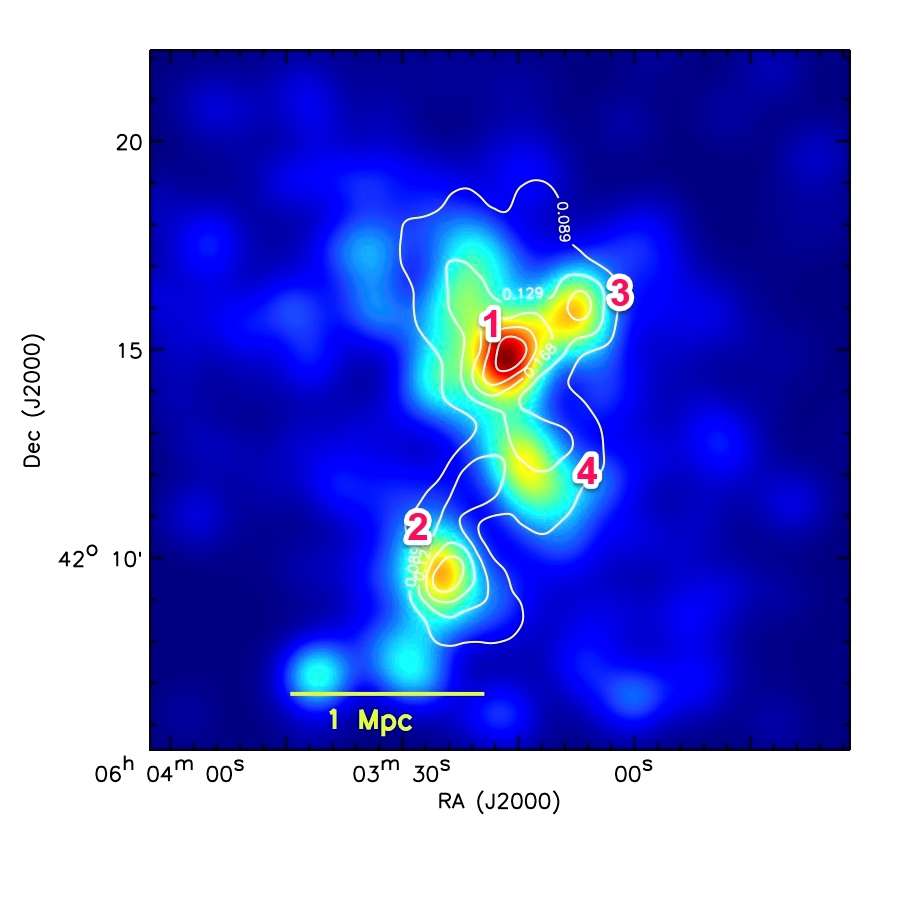}
\caption{Same as Figure~\ref{fig_subaru_massmap} except that mass reconstruction is based on both HST and Subaru imaging. The mass map from this joint analysis is consistent with the Subaru-only result while revealing higher-resolution distributions where HST data are available. Note the improved agreement between mass and optical light compared to the Subaru-only case (Figure~\ref{fig_subaru_massmap}).
}
\label{fig_hst_subaru_massmap}
\end{figure*}

We also merge the {\it HST} and Subaru catalogs and perform mass reconstruction over the large Subaru area. We take into account the source redshift difference, although the difference is minor (0.672 vs. 0.697).
The mass reconstruction based on this joint source catalog is presented in Figure~\ref{fig_hst_subaru_massmap}. The mass map from this joint analysis is very similar to the Subaru-only result while revealing more detailed substructures where {\it HST} data are available.
The centroids of P2 and P4 between the two mass reconstructions agree within $\mytilde0\farcm2$, which is expected from the noise level
of the Subaru-only result. It is not trivial to compare the centroids of P1 and P3 because the two mass peaks are not resolved in the Subaru-only version. Nevertheless, when we smooth the mass map from the HST-Subaru joint analysis, the centroid of the merged
mass peak differs from that of the Subaru-only version by $\mytilde0\farcm5$, which is again consistent with the noise\footnote{We estimate the 
centroid errors by utilizing the publicly available {\tt FIATMAP} code (Fischer \& Tyson 1997), which, although producing a noisier mass map than our maximum-entropy result, runs faster a few hundred times.}.

The significance of the mass substructures in \tb~can be estimated by measuring the background level $\kappa_{bg}$ and rms fluctuation value $\sigma_{\kappa}$ at the substructure locations. That is, we measure the significance $\nu$ via the following:
\begin{equation}
\nu = \frac{\kappa - \kappa_{bg}}{\sigma_{\kappa}}.
\end{equation}
We consider two methods for estimating $\sigma_{\kappa}$. One method is to carry out bootstrapping reanalysis by randomizing source galaxies. Then, it is straightforward to estimate $\sigma_{\kappa}$ from these many mass map realizations.
The other method is to employ the Hessian matrix. The elements of the Hessian matrix are the second derivatives of the likelihood function. Under the assumption that the error distribution at the peak of the posterior distribution is Gaussian, we can convert the Hessian matrix elements into the uncertainties of $\kappa$ at each mass pixel. Because our mass reconstruction based on the maximum entropy algorithm requires considerable CPU time, it is convenient to use the second method in our study. For the evaluation of the background level, we analyze the $\kappa$ values in the four $d=80\arcsec$ strips along the edges of the mass map and
take the 3-$\sigma$-clipped median.

Table~2 displays the significance values of the four mass peaks (shown in Figure~\ref{fig_hst_subaru_massmap}) utilizing the Hessian matrix.
Our significance estimation shows that the two dominant mass peaks (P1 and P2) are solid, corresponding 4.39 $\sigma$ and 3.95 $\sigma$, respectively. The significance estimates of P3 and P4 are somewhat weaker than these two peaks, but their detections are statistically nonnegligible.

\begin{deluxetable*}{ccccc}
\tabletypesize{\scriptsize}
\tablecaption{Significance of substructures in \tb}
\tablenum{2}
\tablehead{\colhead{Property} &  \colhead{P1} & \colhead{P2} & \colhead{P3} & \colhead{P4} }
\tablewidth{0pt}
\startdata
Centroid ($\alpha,\delta$)  & $(90\fdg81674, 42\fdg24823)$   &  $(90\fdg85078, 42\fdg15961)$ & $(90\fdg77961, 42\fdg26730)$ & $(90\fdg79856, 42\fdg21122)$ \\
Significance ($\sigma$)  & 4.39 &  3.95    &  3.08   &  2.60   
\enddata
\tablecomments{When measuring the significance of each peak, we use an aperture whose radius is 325 kpc ($\mytilde90\arcsec$). This value is
approximately the maximum radius, which can prevent the four circular apertures from overlapping with one another. Significance is
estimated by dividing the background-subtracted mean surface density by the mean rms within the aperture. For evaluation of rms, we employ the Hessian matrix while assuming that the error distribution is Gaussian. The background level is calculated by analyzing the four $d=80\arcsec$ strips along the edges of the mass map.}
\end{deluxetable*}

One potentially interesting feature that appears in this joint analysis map, but not seen in the Subaru-only mass map (Figure~\ref{fig_hst_massmap}), is an overdensity near the southern X-ray peak, whose significance is about $2.8\sigma$
We find no luminous cluster members in this region, which is reminiscent
of the ``dark core'' in A520 (Jee et al. 2014; 2012; Mahdavi et al. 2007). However, robust interpretation is difficult without full {\it HST} coverage in this area;
in addition, a significant area in this region is also affected by the bright star mentioned above.

With this caveat, the result shows that the mass structure of \tb~is by and large
bimodal with the two mass components corresponding to the two strongest luminosity peaks. These two mass peaks are also collinear with the two X-ray peaks and the ``brush" of the ``Toothbrush" radio relic.

\subsection{Mass Estimation}
Accurate determination of cluster masses is challenging even for relaxed systems. 
Various systematic errors, as well as differences in analysis method, can lead to $\mytilde10$\%
offsets for the population mean and up to $\mytilde50$\% scatters for individual systems among different 
studies (e.g., von der Linden et al. 2014; Merten et al. 2015; Hoekstra et al. 2015 ).

The issue becomes even more complex when one studies merging clusters. To say the least, merging clusters are believed to show  more departure from conventional analytic profiles, and their lensing signals should be modeled as a superposition of a few or more halos. Often, determining how many halos should be assumed and where they are placed for a given system is not straightforward. In Jee et al. (2014; 2015), we demonstrated that catastrophic ($\gtrsim50$\%) over- or under-estimation can arise if the traditional method, which applies a single analytic profile to azimuthally averaged lensing signals, is employed.

In mass estimation of \tb, we follow the approach of Jee et al. (2014; 2015), where the merging systems are modeled as a binary system. 
This binary assumption can
be considered questionable in~\tb, where the cluster galaxy and mass distributions are somewhat complex, but the inclusion of more than two halos leads to numerical instability in the current case.
Nevertheless, since our final mass map based on {\it HST} and Subaru indicates that the total mass is dominated by the two strongest mass peaks associated with the two 
most luminous halos, we believe that the amount of bias with a two-halo model would not be substantial.
In \textsection\ref{section_mass_left_behind}, we demonstrate that this bias, if any, is
indeed small and within statistical errors by comparing this mass estimate with aperture mass densitometry, which does not require assumptions on the underlying mass distribution.

While fitting two NFW profiles simultaneously, we assume the mass-concentration relation of Duffy et al. (2008) and fix the two halo centers 
at the two brightest luminosity centers. Unlike \ciza, fixing the centers is necessary for \tb~ because of
the relatively low mass (thus low amplitude of the lensing signal) of the system.
We excluded the shears at the core of each halo by choosing $r_{min}=200\arcsec$ and $=50\arcsec$ for P1 and P2, respectively in order to avoid potential bias from enhanced cluster member contamination near the core and signal confusion between strong/weak lensing regimes. The Duffy et al. (2008) relation has a considerable scatter. In Jee et al. (2009), we examined the impact of this scatter on the mass uncertainty of the high-redshift cluster XMMU J2235.3-2557 at $z=1.4$ and showed that the uncertainty from this mass-concentration scatter is $\mytilde11$\%. As this error from the mass-concentration scatter is independent of the statistical noise set by the finite number of source galaxies, in principle the total uncertainty can be obtained by adding the two uncertainties in quadrature. In the current study, we omit the procedure because the total error budget is dominated by the statistical errors.

The resulting $M_{200}$ values for the northern and southern halos are $M_{200}=6.29_{-1.62}^{+2.24}\times10^{14}  M_{\sun}$ and $1.98_{-0.74}^{+1.24} \times10^{14}  M_{\sun}$, respectively (Table 3).
This shows that \tb~consists of two subclusters with an approximate mass ratio of 3:1. 
In merging clusters, estimation of the total mass of the entire system (e.g., $M_{200}$ when the two halos are combined) by adding the masses of the two halos is ambiguous because the result certainly depends on the choice of the system center.
If we choose the geometric mean of the two halos as the center, the total mass of the \tb~system becomes $M_{200}=9.6_{-1.5}^{+2.1} \times10^{14}  M_{\sun}$; 
we determined the value $r_{200}$ numerically by overlapping the two halos in 3D.
This mass nicely agrees with the value $M_{200}=10.1_{-1.4}^{+1.6}\times10^{14} M_{\sun}$ obtained from the tangential shear fitting discussed in \textsection\ref{section_1D} (i.e., assuming
a single halo).
Normally, this level of agreement should be considered surprising
in merging clusters. However, because the southern cluster's contribution to the total mass
is small ($\mytilde2\times10^{14}M_{\sun}$), this agreement is not totally unexpected in \tb. 
We summarize the mass estimation results in Table 3.

\begin{deluxetable*}{ccc}
\tabletypesize{\scriptsize}
\tablecaption{Weak-lensing mass estimation of \tb}
\tablenum{3}
\tablehead{\colhead{Component} & \colhead{Concentration} & 
           \colhead{$M_{200}~(\times10^{14}  M_{\sun}$)}  }
\tablewidth{0pt}
\startdata
North   &  $3.30\pm0.08$ & $6.29_{-1.62}^{+2.24}$  \\
South   &   $3.64\pm0.14$ & $1.98_{-0.74}^{+1.24}$  \\
Total (two-component)$^1$  &        -         & $9.6_{-1.5}^{+2.1}$ \\
Total (one-component)$^2$  &   3.17$\pm$0.04         & $10.1_{-1.4}^{+1.6} $
\enddata
\tablecomments{1. We compute the total mass of the system by adopting the geometric center of the two components as the center and estimating the combined mass (superposition of two halos) within $r_{200}$, inside
which the mean density becomes 200 times the critical density at $z=0.225$.
2. We use the tangential shear profile at $r>200\arcsec$ to estimate the total mass. }
\end{deluxetable*}

\section{DISCUSSION} \label{section_discussion}

\subsection{Any Mass Left Behind?} \label{section_mass_left_behind}

Because we make an approximation that the mass of \tb~is dominated by the two halos associated with the two X-ray peaks, it is useful to examine the validity of the assumption by an independent method. We employ aperture mass densitometry, which allows us to estimate total projected masses within a given aperture without any assumption on the number of halos and their profiles.
We will compare this projected mass from aperture mass densitometry with the results from our two halo model by projecting the 3D NFW mass distribution onto the plane of sky.

Aperture mass densitometry (Fahlman et al. 1994; Clowe et al. 2000) is computed through the following equation:
\begin{eqnarray}
\zeta _c (r_1, r_2,r_{max})  =  \bar{\kappa}( r \leq r_1) -
\bar{\kappa}( r_2 < r \leq r_{max}) \nonumber \\ =  2 \int_{r_1} ^{r_2} \frac{
  \left < \gamma_T \right > }{r}d r + \frac{2}{1-r_2^2/r_{max}^2}
\int_{r_2}^{r_{max}} \frac{ \left <\gamma_T \right >}{r} d r,
\label{eqn_aperture_densitometry}
\end{eqnarray}
\noindent
where $\langle \gamma_T \rangle$ is the azimuthal average of
tangential shears, $r_1$ is the aperture radius, and $r_2$ and
$r_{max}$ are the inner and the outer radii of the annulus, respectively.
$\zeta_c(r_1,r_2,r_{max})$
provides a density contrast of the region inside $r<r_1$ with respect
to the control annulus $(r_2,r_{max})$. We choose $r_2=800\arcsec (\mytilde2.9~\mbox{Mpc})$ and
$r_{max}=1000\arcsec (\mytilde3.6~\mbox{Mpc})$ for the control annulus. 
Projecting our NFW fitting results, we estimate the density
within this annulus to be $\bar{\kappa}=0.004$.
Because the control annulus radius is large and the density there is  small, the impact of adopting the NFW results on the aperture mass densitometry becomes negligible.

\begin{figure}
\includegraphics[width=8.5cm,trim=0cm 0cm 0cm 0cm]{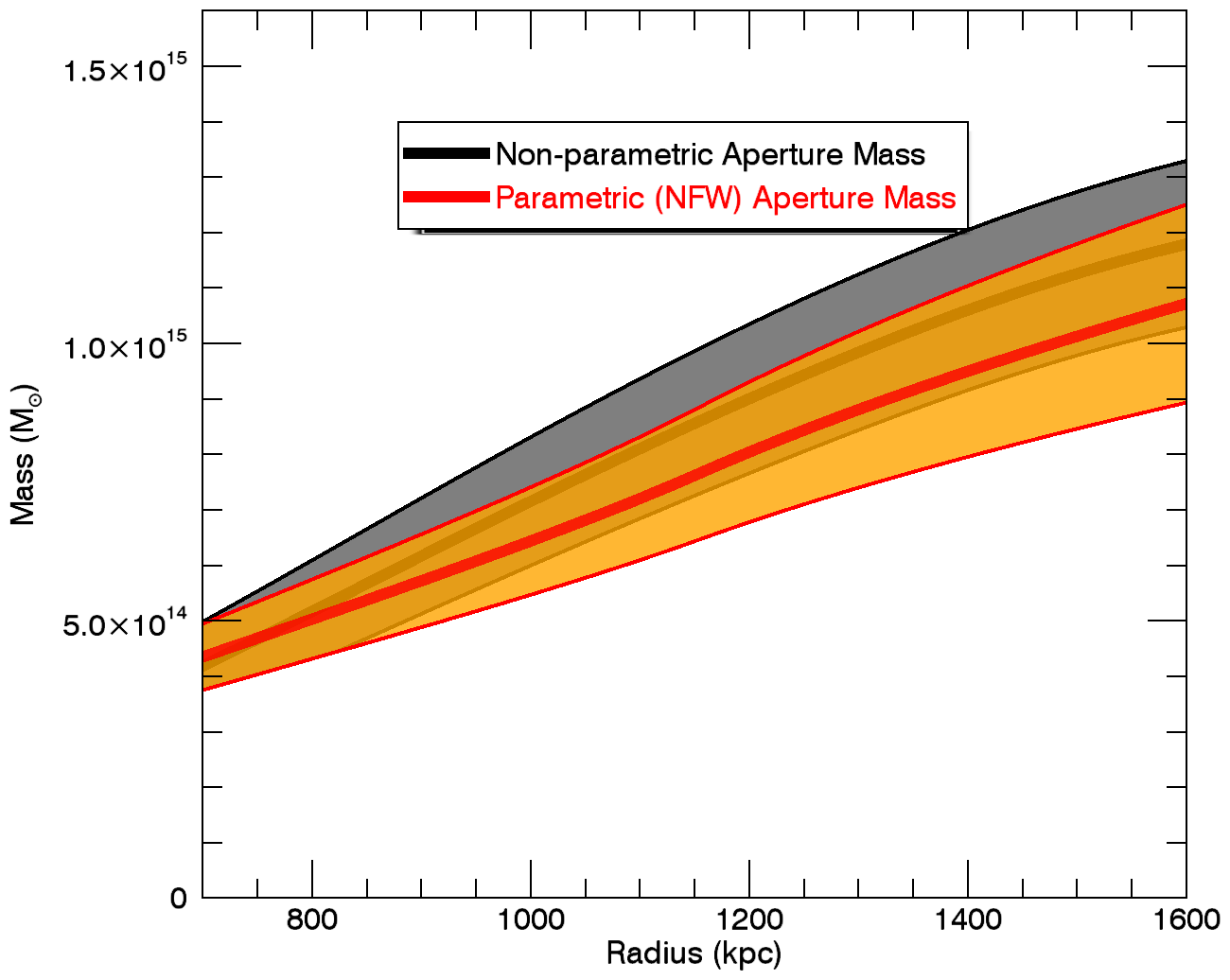}
\caption{Projected masses of \tb. We compare nonparametric (aperture mass densitometry) results with parametric ones (analytic projection of two NFW profiles). The band of each color shows the 1-$\sigma$ range of statistical uncertainties. 
The two results are consistent with each other, and we conclude that no significant mass is excluded by modeling \tb~with two NFW profiles. 
}
\label{fig_aperture_mass}
\end{figure}

The input to the equation of the densitometry is a shear $\gamma_T$, not a reduced shear $g_T$. Therefore, we need to determine the aperture mass using the relation $\gamma=(1-\kappa)g$. We find that the density converges after three or four iterations.
The resulting aperture mass is displayed in Figure~\ref{fig_aperture_mass}. Also displayed in Figure~\ref{fig_aperture_mass} is the aperture mass estimated by projecting the NFW fitting results above. In order to obtain this estimation, we first projected each NFW profile along the line-of-sight direction and summed the two-halo results. 
The aperture mass density masses are within the 1$\sigma$ upper limits of the NFW masses, which may hint at the possibility that the two-halo representation may not be a perfect choice. 
However, because the 1$\sigma$ error bars from both methods overlap, we argue that the difference should not be considered statistically significant.

\subsection{Comparison with X-ray Results and Implication for the ``Toothbrush" Merging Scenario}

A diffuse hot plasma within a cluster is well traced by X-ray emission because the  emissivity is in general
proportional to the plasma density squared (given the same plasma temperature). Since the plasma consists of charged particles subject to ram pressure, the X-ray morphology of merging clusters reveals critical information that cannot be probed otherwise.
Here we compare the X-ray morphology of \tb~with the weak-lensing mass distribution and discuss the implication in the context of the merging scenario responsible for the observed ``Toothbrush" radio relic.

\tb~has been observed with both {\it XMM-Newton} and {\it Chandra}. The 82~ks {\it XMM-Newton} data were studied by Ogrean et al. (2013), and van Weeren et al. (2015) analyzed the 237 ks $Chandra$ data ({\tt ObsID}: 15171, 15172, and 15323). Ogrean et al. (2013) showed that the intracluster medium of \tb~is dominated by two components, which is confirmed by the Chandra study of van Weeren et al. (2015). In addition, a few new remarkable features are revealed in the high-resolution Chandra observation. First, a density jump indicating a shock is
detected in the southern edge. This location coincides with the southern edge of the radio halo. Across the shock a temperature jump is also found. The two Mach numbers derived by both density and temperature jumps
are consistent ($\mathcal{M}=1.4_{-0.058}^{+0.063}$ and $1.7_{-0.3}^{+0.5}$, respectively).
Second, the high-resolution Chandra data show that the southern X-ray component has a triangular ``bullet"-like shape. According to their further temperature analysis, the density jump at the southern edge of the bullet indicates a cold front.

\begin{figure}
\includegraphics[width=9cm,trim=1cm 0cm 0cm 0cm]{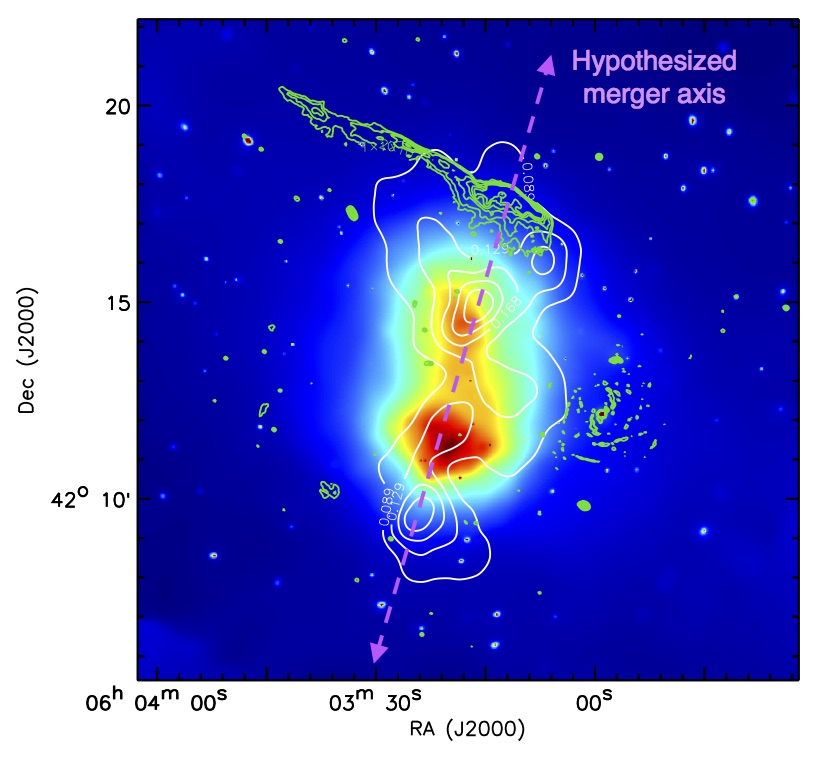}
\caption{Comparison of weak-lensing mass, X-ray, and radio emission. The white contours are mass density. The background is color-coded with the intensity of adaptively smoothed X-ray emission. The green
contours represent the 610 MHz radio (GMRT) intensity.
The northern mass peak is more massive than the southern mass peak approximately by a factor of three, whereas the X-ray emission is much stronger 
in the southern peak. Both X-ray peaks are offset from the corresponding mass peaks. The larger offset between the southern mass and the ``bullet"-like X-ray peak is consistent with our expectation because the smaller mass 
must experience larger ram pressure. Note the  collinearity of the two X-ray peaks, two strongest mass peaks, and the ``brush" of the ``Toothbrush" relic, which we hypothesize as the merger axis.
}
\label{fig_massoverxray}
\end{figure}

The comparison of these X-ray findings with the current weak-lensing results provides a consistent picture regarding the merging scenario of \tb. We display
the comparison in Figure~\ref{fig_massoverxray}, where we illustrate our hypothesized merger axis.
Despite the somewhat complex galaxy distribution, the X-ray and weak-lensing results suggest that the dominant merger may be approximated by a north-south collision of two components.
We find offsets between X-ray and mass peaks. The northern mass peak is offset toward the northwest with respect
to the corresponding X-ray peak by $\mytilde0.5\arcmin$ whereas the the southern mass peak is offset to the south by $\mytilde2\arcmin$. Similar to the Bullet Cluster, the direction of the offsets favors a scenario wherein the two components passed through each other and are still separating.
Our weak-lensing analysis shows that the northern component is more massive than the southern component by a factor of three. We believe that this mass inequality is consistent with the offset inequality, since the less massive southern component should experience more ram pressure. 
Another supporting evidence for this mass inequality is the location of the ``Toothbrush"-relic. The simulation by van Weeren et al. (2011) predicts that two radio-relics are generated in a two-body encounter and travel along the merger axis with
the larger relic associated with the more massive halo.
The observation that the $\mytilde2~$Mpc ``Toothbrush"-relic is located near the northern edge of \tb~ is consistent with the northern component being more massive in our weak-lensing analysis. The same trend has been observed in our weak-lensing study of the ``sausage" cluster \ciza~(Jee et al. 2015) and ZwCl0008.8+5215 (N. Golovich et al. in prep.). 
The study of ZwCl0008.8+5215 shows that the X-ray emission of the less massive system appears as a clear ``bullet"-like feature, similar to the ``Bullet" cluster (Clowe et al. 2006), whereas the larger radio relic is found near the edge of the more massive system.
The exact physical mechanism is unknown as to the question ``why does the larger radio relic occur on the higher mass side?" On the other hand, X-ray observations show that distinct shock features such as density discontinuities, temperature jumps, etc., are more prominent on the lower-mass side unlike radio relics, which are also believed to trace the location of shock fronts.

Some may argue that the above mass inequality argument may be challenged by the X-ray luminosity of the southern peak
being much higher. Needless to say, in general X-ray luminosity is positively correlated with mass. However, in active merging clusters, it is natural to suspect that this correlation between mass and X-ray luminosity can temporarily be altered for many reasons (e.g., Randall et al. 2002; Skillman et al. 2013). For example, a cool core (associated with a lower-mass component) can survive a head-on collision, whereas a hot core (associated with a 
higher-mass component) can severely be disrupted after a core pass-through. The deep Chandra X-ray image of the ``El Gordo" cluster at $z=0.87$ is a good example. The weak-lensing study shows that the system is composed of two halos with a 2:1 mass ratio, whereas only the cool core associated with the less massive system (south) is clearly visible in X-ray (Jee et al. 2014; Menanteau et al. 2012).
The hydrodynamical simulation by Molnar \& Broadhurst (2015) reproduces this asymmetry in brightness between the two
X-ray peaks of ``El Gordo".
Another example is ZwCl0008.8+5215 (N. Golovich et al. in prep.) at $z=0.1$ mentioned above. By and large the ZwCl0008.8+5215 cluster is also a binary merging system with one of two X-ray peaks resembling a ``bullet"-like shape. The ``bullet" component is brighter than the other component in X-ray, whereas the mass associated with the ``bullet" is found to be approximately a factor of five smaller.

\subsection{Remaining Puzzles of the ``Toothbrush"-relic Cluster}
One of the goals of the MC$^2$ collaboration is to enable
quantitative comparisons between observations and simulations for interesting merging clusters. \tb~is a remarkably interesting system and should be followed up by careful numerical analysis. Here we discuss some of the puzzling issues of \tb~that future hydrodynamic simulations should address.

First, we believe that the extremely high ICM temperature of \tb~deserves our attention.
Van Weeren et al. (2015) 
constrain the temperatures of the northern and southern X-ray peaks to be $8.43_{-0.25}^{+0.26}$ keV and $9.00\pm0.28$ keV, respectively. These temperatures are substantially higher than what our weak-lensing masses imply {\it if} we neglect nonthermal energy injection into ICM.
With the assumption of the isothermal $\beta$ model with $r_c=100$ kpc and $\beta_X=0.7$, the weak-lensing masses are  converted to $T_X\sim5$ keV and $\sim2$ keV for
the northern and southern halos, respectively. 
Even when we treat \tb~ as a single halo with $M_{200}\sim9.6\times10^{15} M_{\sun}$ (i.e., the sum of the two halos using the values in Table 3), the implied temperature (again with an isothermal $\beta$ model) becomes only $T_X\sim7$ keV, significantly smaller than the global X-ray temperature $T_X\sim10$ keV.
Although this discrepancy may not be considered surprising, given the common understanding that X-ray temperatures of merging clusters are biased indicators of the cluster masses, the level of discrepancy that we witness in \tb~ is somewhat extreme when we consider results for other clusters in the literature. For example,
even for the ``Bullet"-cluster, Clowe et al. (2006) find that the temperature levels of the system are consistent with their weak-lensing masses.

Second, although we argue that the two subclusters of \tb~ played the dominant roles in creating the current observational features such as the galaxy-mass-relic alignments, the offsets between mass/galaxy and X-ray peaks, etc., the long asymmetrically linear feature of the ``Toothbrush" relic strongly suggests that a smaller third component might have been
involved, as suggested by Bruggen et al. (2012).
However, the mass, path, and timing of this third component are unclear. It is our hope that the weak-lensing 
substructures revealed in the current study will aid us to reduce the volume of the parameter space that future simulations should explore.

Third, the implied collision velocity is very high. The high polarization fraction $\alpha\lesssim60$\% (van Weeren et al. 2012) suggests that the merger may be happening in the plane of the sky. According to our redshift analysis (W. Dawson et al. in prep), the line-of-sight velocity difference between the northern and southern subclusters is $\mytilde1800~$km~s$^{-1}$.
Even with the polarization prior $\alpha\sim30$\%, the implied collision velocity is as high as $\mytilde$3500~km~s$^{-1}$, which exceeds the escape velocity of the \tb~system and thus is hard to accommodate within the current $\Lambda$CDM paradigm. More detailed discussions will appear in W. Dawson et al. (in prep).

\section{CONCLUSIONS} \label{section_conclusion}

We have presented detailed weak-lensing studies of the ``Toothbrush" relic cluster \tb~with {\it HST} and Subaru imaging.
Together with the ``Sausage" relic cluster \ciza, \tb~ has been known for its giant ($\mytilde2$Mpc) radio relic, whose peculiar morphology gives the system the nickname ``Toothbrush". 

Our weak-lensing study provides a high-resolution map of the cluster dark matter, which helps
us to constrain the merging scenario responsible for the ``Toothbrush" relic. We find that although the cluster substructure is more complicated than that of \ciza, the global mass distribution can be approximated by a bimodal distribution with a 3:1 mass ratio.
The northern mass clump encloses $M_{200}=6.29_{-1.62}^{+2.24}\times 10^{14} M_{\sun}$ and coincides with the galaxy luminosity peak.
The southern mass component contains $M_{200}=1.98_{-0.74}^{+1.24} \times10^{14} M_{\sun}$ and is also in an excellent spatial agreement with the southern luminosity peak. However, the southern mass peak is $\mytilde2\arcmin$ offset with respect to the southern X-ray peak.
The two mass peaks, two X-ray peaks, two luminosity peaks, and the
``brush" of the ``Toothbrush" relic are collinear, which strongly suggests that the violent merger responsible for the giant radio relic was mainly derived by the collision of the two components. However, the long ``handle" relic extended northeast from the
``brush" indicates that a third (or more) component might have been involved in this merger.
It is interesting that our weak-lensing mass reconstruction reveals a mass clump southwest
of the northern mass peak. We find that a galaxy luminosity peak coincides with this 
mass overdensity. Nevertheless, we have yet to carry out detailed simulations in order to quantify the possibility that this observed component might have been involved in the creation of the peculiar radio-relic morphology.

The shape of the southern X-ray peak is triangular and is reminiscent of
the ``Bullet" in the Bullet Cluster. A recent Chandra study reveals a shock south of
this feature traced by both density and temperature jumps. Together with the aforementioned offset, these X-ray features show that we may be witnessing a post-collision, where the two cluster components are separating from each other.

The high X-ray temperatures of \tb~are  discrepant with what the weak-lensing masses imply. We attribute the large differences to substantial departure from the hydrostatic equilibrium.
These severe discrepancies support the consensus that using X-ray 
temperatures is an unreliable way to infer cluster masses in violent merging systems.

Galaxy clusters are receiving growing attention as cosmic particle accelerators. Although every merger case is special and deserves scrutiny, radio-relic clusters are particularly useful thanks to strong constraints on both the geometry and stage of the mergers, which enables us to reduce the parameter search space by substantial factors. Of course, careful numerical simulations should follow up the observations in order to come up with quantitatively coherent scenarios, wherein all the observed features fit together within the observational uncertainties.\\

Support for Program number HST-GO-13343.01-A
was provided by NASA through a grant from the Space Telescope Science
Institute, which is operated by the Association of Universities for
Research in Astronomy, Incorporated, under NASA contract NAS5-26555. MJJ acknowledges support from NRF of Korea to CGER.
AS acknowledges financial support from an NWO top subsidy (614.001.006).
HR gratefully acknowledges support from the European Research Council under the European Union’s Seventh Framework Programme (FP/2007-2013)/ERC Advanced Grant NewClusters-321271.
Part of this work performed under the auspices of the U.S. DOE by LLNL under Contract DE-AC52-07NA27344

\end{document}